\documentclass[aps,prl,superscriptaddress,twocolumn,showpacs]{revtex4-1}
\usepackage{hyperref}
\usepackage{graphicx,epsfig}
\usepackage{subeqn}
\usepackage[usenames]{color}

\newcommand{\bk}{{\bf k}}
\newcommand{\bl}{{\bf l}}

\newcommand{\cH}{{\mathcal H}}

\newcommand{\cZ}{{\mathcal Z}}

\begin{document}

\title{Multiple solutions for the equilibrium populations in BCS superconductors}

\author{Drago\c s-Victor Anghel}
\email{dragos@theory.nipne.ro}
\affiliation{Horia Hulubei National Institute for Physics and Nuclear Engineering, P.O. Box MG-6, 077126 M\u agurele, Ilfov, Romania}

\date{\today}

\begin{abstract}
It was recently shown that the BCS formalism leads to several solutions for the energy gap and the equilibrium quasiparticle distribution, with a phase transition temperature which depends on the position of the chemical potential within the attraction band (the attraction band AB is defined as the single-particle energy interval in which the pairing interaction is manifested). Moreover, in some cases, the phase transition may be of the first, not of the second order. Here I will find two sets of solutions for any temperature below the phase transition temperature. I will also show that, when the AB is symmetric with respect to the chemical potential (the textbook BCS problem) there are still two solutions, with different energy gaps: one solution is the typical (textbook) BCS solution, whereas the other one has a smaller energy gap and non-zero quasiparticle populations down to zero temperature. At zero temperature, the energy gap corresponding to the second solution is one third of the typical BCS solution.
% 
% It was recently shown that the BCS formalism leads to several solutions for the energy gap and the equilibrium quasiparticle distribution.
% While this became quite obvious when the attraction band (which is the single-particle energy interval in which the pairing interaction is manifested) is asymmetric with respect to the chemical potential of the system (that is, the center of the attraction band $\mu$ is different from the chemical potential $\mu_R$), I will show here that there are two solutions, with different energy gaps and quasiparticle populations even when $\mu = \mu_R$.
% One of the solution leads to the well-known BCS results, with an energy gap denoted here by $\Delta_1(\mu_R-\mu = 0, T)$ (where $T$ is the temperature), whereas the second one leads to an energy gap $\Delta_2(\mu_R-\mu = 0, T) \le \Delta_1(\mu_R-\mu = 0, T)$, for any $T$ below the critical temperature -- the critical temperature is the same for both solutions.
% In the second solution, the quasiparticle population is different from zero even at $T=0$ and it was shown before that $\Delta_2(\mu_R-\mu = 0, T=0) = \Delta_1(\mu_R-\mu = 0, T=0)/3$.
\end{abstract}

\maketitle

\section{Introduction} \label{sec_intro}

The achievement of room temperature superconductivity is one of the major goals of science, with vast technological implications.
The progress in this direction has been slow for many decades, from the discovery of superconductivity, in 1911~\cite{CommPhysLabUnivLeiden.120b.1911.Onnes, CommPhysLabUnivLeiden.122b.1911.Onnes, CommPhysLabUnivLeiden.124c.1911.Onnes},  until the discovery of superconductivity in cuprates, in 1986~\cite{ZPhysB.64.189.1986.Bednorz}.
The discovery of Bednorz and M\"uller~\cite{ZPhysB.64.189.1986.Bednorz} marked the beginning of the high temperature (or high-$T_c$) superconductivity, with the critical temperature $T_c$ rising fast, from approximately 30~K (at that time, this was  believed to be the highest possible transition temperature achievable by BCS superconductors) to 133~K, in 1993, in the Hg-Ba-Ca-Cu-O compound~\cite{Nature.363.56.1993.Schilling}.
One year later, maximum critical temperature $T_c \approx 164$~K was eventually achieved in cuprates, also in Hg-Ba-Ca-Cu-O compounds, but under high pressure (up to 45~GPa)~\cite{PhysRevB.50.4260.1994.Gao}.

The discovery of superconductivity in iron-based materials (so called iron-superconductors)~\cite{JAmChemSoc.128.10012.2006.Kamihara} and the observation that they may have high critical temperatures~\cite{JAmChemSoc.130.3296.2008.Kamihara, Nature.453.376.2008.Takahashi} produced a new  wave of high interest in room temperature superconductivity.
Eventually, the highest critical temperatures obtained with iron-based superconductors was above 100~K~\cite{ChinPhysLett.29.037402.2010.Wang, NatComm.3.931.2012.Liu, NatMat.12.605.2013.He, NatMat.14.285.2015.Ge}, but that happened in ultra-thin films and is still far from reaching the room temperature superconductivity goal.

Beside changing the chemical composition and applying pressure, the superconducting properties of materials can be altered by applying external electromagnetic waves~\cite{Science.331.189.2011.Fausti, NatMater.13.705.2014.Hu, PhysRevB.89.184516.2014.Kaiser, Nature.530.461.2016.Mitrano, PhysRevB.98.144513.2018.Secchi}, but the results are far from breaking the previously established records (mentioned above).
% Another method to modify the superconducting properties of materials is to apply external electromagnetic waves~\cite{Science.331.189.2011.Fausti, NatMater.13.705.2014.Hu, PhysRevB.89.184516.2014.Kaiser, Nature.530.461.2016.Mitrano, PhysRevB.98.144513.2018.Secchi}.

The progress in unconventional superconductivity (i.e. superconductivity in cuprates and iron-based materials) is hampered by the fact that the microscopic mechanisms responsible for it are not yet known, which makes theoretical predictions very difficult.
% For this reason, theoretical predictions are hard to make.
Without consistent theoretical guidance, the search for the optimal materials and experimental conditions is hard.
From this point of view, the situation for conventional superconductors (also called metallic or BCS superconductors) is much better.
Old theoretical predictions by made N. W. Ashcroft suggest that high temperature superconductivity may be achieved in metallic hydrogen~\cite{PhysRevLett.21.1748.1968.Ashcroft} or in hydrogen dominant metallic alloys~\cite{PhysRevLett.92.187002.2004.Ashcroft}.
These predictions have been put to test and in a relatively short interval of time considerable progress has been made (both, theoretically and experimentally)~\cite{ProcNatlAcadSciUSA.109.6463.2012.Wang, Nature.525.73.2015.Drozdov, ProcNatlAcadSciUSA.114.6990.2017.Liu, PhysRevLett.119.107001.2017.Peng, PhysRevB.98.100102.2018.Liu, PhysRevB.99.140501.2019.Liu, PhysRevLett.122.027001.2019.Somayazulu, NanoLett.19.2537.2019.Xia}, reaching the record critical temperature of 250~K in lanthanum hydride under high pressure~\cite{Nature.569.528.2019.Drozdov}.
Furthermore, very recently, metallic hydrogen has eventually been obtained~\cite{arXiv190605634.Loubeyre}, which increases hopes that room temperature superconductivity will soon be realized.

The BCS theory is at the heart of this tremendous recent progress, but this theory has also been successfully applied to other fields of physics, like nuclear physics~\cite{PhysRev.110.936.1958.Bohr}, astrophysics~\cite{PhysRevLett.21.1748.1968.Ashcroft, RevModPhys.75.607.2003.Dean, JAstrophysAstr.2017.43.2017.Chamel}, etc.
Considering the extreme conditions in which some of  these systems exist, not only the stable equilibrium configurations predicted by BCS should be considered, but also the metastable ones and, as it has been recently shown, its basic system of equations do not have unique solutions for the energy gap and quasiparticle populations~\cite{PhysicaA.464.74.2016.Anghel, PhysicaA.531.121804.2019.Anghel}.
In Ref.~\cite{PhysicaA.464.74.2016.Anghel} it was shown, for example, that when the attraction band (AB) is asymmetric with respect to the chemical potential (the AB is the single-particle energy interval in which the pairing interaction is manifested),
% the BCS formalism leads to some surprising results.
% In the grandcanonical ensemble,
some solutions are different from the standard BCS solutions, with different energy gaps and asymmetric quasiparticle populations (the asymmetry persists even at zero temperature).
The phase transition temperatures of the new solutions decrease with the asymmetry of the AB.
Furthermore, if the chemical potential is fixed and charging effects are not taken into account (i.e., the  grandcanonical ensemble is used) the phase transition is accompanied by a change in the total number of particles in the system.

The results of Ref.~\cite{PhysicaA.464.74.2016.Anghel} were analyzed in more detail at zero temperature in Ref.~\cite{PhysicaA.531.121804.2019.Anghel} and it was observed that the new method leads to two families of results -- that is, for each asymmetry parameter there are two solutions for the gap equation and, consequently, for the quasiparticle populations.
Nevertheless, if the conservation of the number of particles is imposed, only one of the solutions remains.

Various levels of asymmetries of the AB in a superconductor may occur not only because of the electronic band structure or the microscopic interactions, but may also be induced by external electromagnetic waves, as has been shown for example in Ref.~\cite{PhysRevB.98.144513.2018.Secchi}.
So, a consistent analysis of the BCS formalism require the study of metastable states in the presence of asymmetric ABs.
For this reason, in this paper I show that both solutions found in Ref.~\cite{PhysicaA.531.121804.2019.Anghel} survive at finite temperatures.
I analyze the variation of the energy gap with the temperature and the asymmetry parameter and I find the critical temperature.
Most interesting, I show that the two solutions survive at finite temperatures even in the limit of symmetric AB.
In other words, the standard BCS formalism, with symmetric AB, admits not only the standard, textbook solution~\cite{Tinkham:book, PhysRev.108.1175.1957.Bardeen}, but also a second solution.

\section{Methods} \label{sec_methods}

We start from the BCS Hamiltonian~\cite{Tinkham:book, PhysicaA.464.74.2016.Anghel},
\begin{equation}
  \hat\cH = \sum_{\bk s}\epsilon^{(0)}_\bk \hat n_{\bk s} + \sum_{\bk\bl} V_{\bk\bl} c^\dagger_{\bk\uparrow} c^\dagger_{-\bk\downarrow} c_{-\bl\downarrow} c_{\bl\uparrow} , \label{def_H_BCS}
\end{equation}
where we denoted by $c^\dagger_{\bk, s}$ and $c_{\bk,s}$ the creation and annihilation operators on the free-particle state $|\bk,s\rangle$, respectively; $\bk$ and $s$ are some quantum numbers, for example wavevector and spin projection ($\uparrow, \downarrow$), respectively.
The energy of the free-particle state is $\epsilon_\bk^{(0)}$, whereas $\hat n_{\bk s} \equiv c^\dagger_{\bk, s} c_{\bk,s}$ is the occupation number operator.
For the convenience of the  calculations one usually assumes that the pairing potential $V_{\bk\bl} \equiv V$ is constant and different from zero if and only if $\epsilon_\bk^{(0)}$ and $\epsilon_\bl^{(0)}$ belong to a finite interval $I_V \equiv [\mu - \hbar\omega_c, \mu + \hbar\omega_c]$ centered at $\mu$.
The model Hamiltonian $\hat\cH_M = \hat\cH - \mu \hat N$ ($\hat N \equiv \sum_{\bk,s} c^\dagger_{\bk s} c_{\bk s}$ is the total particle number operator) may be diagonalized by the Bogoliubov-Valatin transformations~\cite{NuovoCimento.7.843.1958.Valatin, NuovoCimento.7.794.1958.Bogoljubov} to obtain
\begin{equation}
  \hat\cH_M = \sum_\bk(\xi_\bk-\epsilon_\bk+\Delta b_\bk^*) + \sum_\bk\epsilon_\bk(\gamma^\dagger_{\bk 0}\gamma_{\bk 0} + \gamma^\dagger_{\bk 1}\gamma_{\bk 1}) , \label{HM_BCS}
\end{equation}
where $\xi_\bk\equiv \epsilon^{(0)}_\bk - \mu$, $\epsilon_\bk \equiv \sqrt{\xi_\bk^2+\Delta^2}$, $\Delta$ is the \textit{energy gap} defined by the equation
%
% \begin{eqnarray}
  $\Delta \equiv - V \sum_\bl \langle c_{-\bk\downarrow} c_{\bk\uparrow}\rangle$,
%   , \label{def_Delta0}
% \end{eqnarray}
%
whereas $\langle \cdot \rangle$ represents the statistical average.
The operators $\gamma^\dagger_{\bk i}$ and $\gamma_{\bk i}$ ($i=0,1$) -- satisfying
$c_{\bk\uparrow} = u^*_\bk \gamma_{\bk 0} + v_\bk \gamma^\dagger_{\bk 1}$ and $c_{-\bk\downarrow} = -v_\bk \gamma^\dagger_{\bk 0} + u^*_\bk \gamma_{\bk 1}$ --
are the quasiparticle creation and annihilation operators, respectively.
The coefficients $u_\bk$ and $v_\bk$ are
\begin{equation}
  |v_\bk|^2 = 1 - |u_\bk|^2 = \frac{1}{2} \left(1 - \frac{\xi_\bk}{\epsilon_\bk}\right) . \label{def_uv}
\end{equation}
In these notations, the energy gap is obtained by solving self-consistently the equation~\cite{Tinkham:book, PhysicaA.464.74.2016.Anghel},
\begin{subequations} \label{set_eqs1}
\begin{equation}
  1 = \frac{V}{2} \sum_\bk \frac{1 - n_{\bk 0} - n_{\bk 1}}{\epsilon_\bk} . \label{def_Delta2}
\end{equation}
and the quasiparticle populations are given by~\cite{PhysicaA.464.74.2016.Anghel}
%
% \begin{subequations} \label{eqs_pop}
\begin{equation}
  n_{\bk i} \equiv \langle \gamma^\dagger_{\bk i} \gamma_{\bk i} \rangle = \frac{1}{e^{\beta(\epsilon_{\bk}-\tilde\mu_\bk)}+1}, \quad i=0,1 , \label{pop_til_eps}
\end{equation}
where
\begin{equation}
  \tilde\mu_\bk \equiv \frac{\mu_R - \mu}{\epsilon_\bk} \left[ \xi_\bk - \frac{ \sum_\bk \left( 1 - n_{\bk 0} - n_{\bk 1} \right) \xi_\bk \epsilon_\bk^{-3}}
  { \sum_\bk \left(1 - n_{\bk 0} - n_{\bk 1} \right) \epsilon_\bk^{-3} } \right]  \label{def_tilde_mu} \\
  %
%   \tilde\mu \equiv \frac{\mu_R - \mu}{\epsilon_\bk} \left[ \xi_\bk - \frac{ \int_{-\hbar\omega_c}^{\hbar\omega_c} \sigma(\xi+\mu) ( 1 - n_{\xi 0} - n_{\xi 1} ) \frac{\xi}{\epsilon^3} \, d\xi }
%   { \int_{-\hbar\omega_c}^{\hbar\omega_c} \frac{(1 - n_{0\xi} - n_{1\xi}) \sigma(\xi+\mu) d\xi}{\epsilon^3} } \right] . \label{def_tilde_mu}
\end{equation}
% \end{subequations}
\end{subequations}
is a correction to the quasiparticle energy and $\mu_R$ is the chemical potential.
Solving self-consistently the set of Eqs.~(\ref{set_eqs1}), one obtains the energy gap and the quasiparticle populations, as exemplified in~\cite{PhysicaA.464.74.2016.Anghel, PhysicaA.531.121804.2019.Anghel}.

If we work in the quasi-continuous limit, assuming a constant density of states (DOS) along the single-particle axis $\epsilon^{(0)}$, namely $\sigma(\epsilon_\bk^{(0)}) = \sigma_0$, the set~(\ref{set_eqs1}) simplifies considerably and becomes~\cite{PhysicaA.464.74.2016.Anghel, PhysicaA.531.121804.2019.Anghel}
\begin{subequations}\label{set_eqs2}
\begin{eqnarray}
  \frac{2}{\sigma_0V} &=& \int_{-\hbar\omega_c}^{\hbar\omega_c} \frac{1 - 2 n_{\xi}}{\sqrt{\xi^2+\Delta^2}} d\xi ,
  \label{Eq_int_Delta1} \\
  F(\mu_R-\mu, T) &\equiv& \frac{ \int_{-\hbar\omega_c}^{\hbar\omega_c} ( 1 - 2n_{\xi} ) \frac{\xi}{\epsilon^3} \, d\xi }
  { \int_{-\hbar\omega_c}^{\hbar\omega_c} \frac{(1 - 2 n_{\xi}) d\xi}{\epsilon^3} } , \label{def_F_sigma0} \\
  n_{\xi}(\mu_R-\mu, T) &=& \frac{1}{e^{\beta[\epsilon_{\xi}-(\mu_R-\mu)(\xi - F)/\epsilon_\xi]}+1} , \label{pop_til_eps_sigma0}
\end{eqnarray}
\end{subequations}
where we took into account that \textit{in equilibrium} $n_{\xi 0} = n_{\xi 1} \equiv n_\xi$ for any $\xi$.
Equation~(\ref{Eq_int_Delta1}) taken separately, with $n_{\xi} = 0$, gives the standard BCS result for the energy gap at zero temperature, $\Delta_0 \approx 2\hbar\omega_c \exp [-1/(\sigma_0V)]$ (in the weak coupling limit~\cite{Tinkham:book}).
If we set $F \equiv 0$ and $\mu = \mu_R$, the set (\ref{set_eqs2}) has only one solution, strictly decreasing and of class $C^2$ with respect to the temperature~\cite{JMathAnalAppl.383.353.2011.Watanabe, AbstrApplAnal.2013.932085.2013.Watanabe}.

\section{Results} \label{sec_results}

\begin{figure}[t]
  \centering
  \includegraphics[width=6cm,bb=0 0 557 482,keepaspectratio=true]{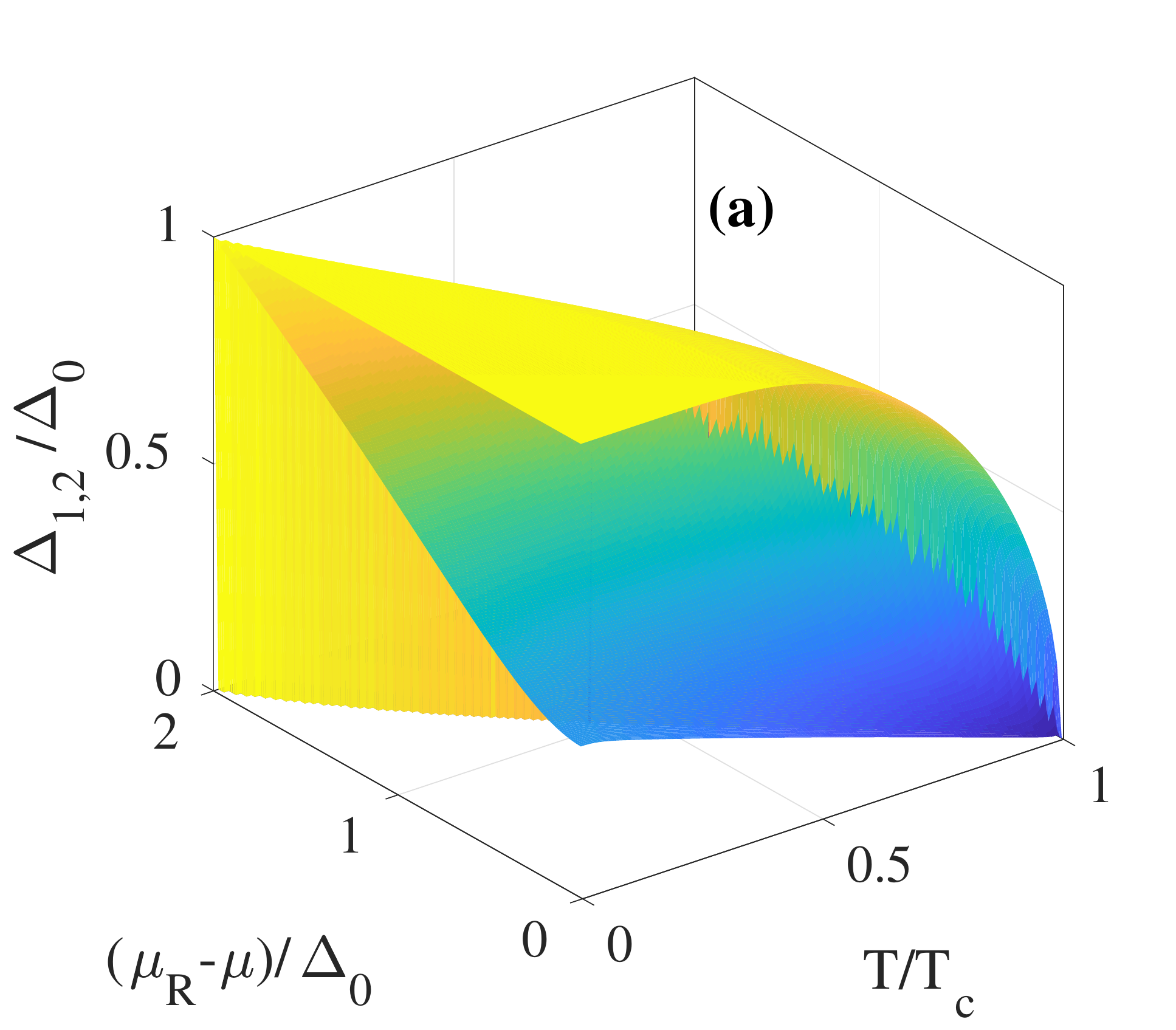}
  \includegraphics[width=6cm,bb=0 0 557 482,keepaspectratio=true]{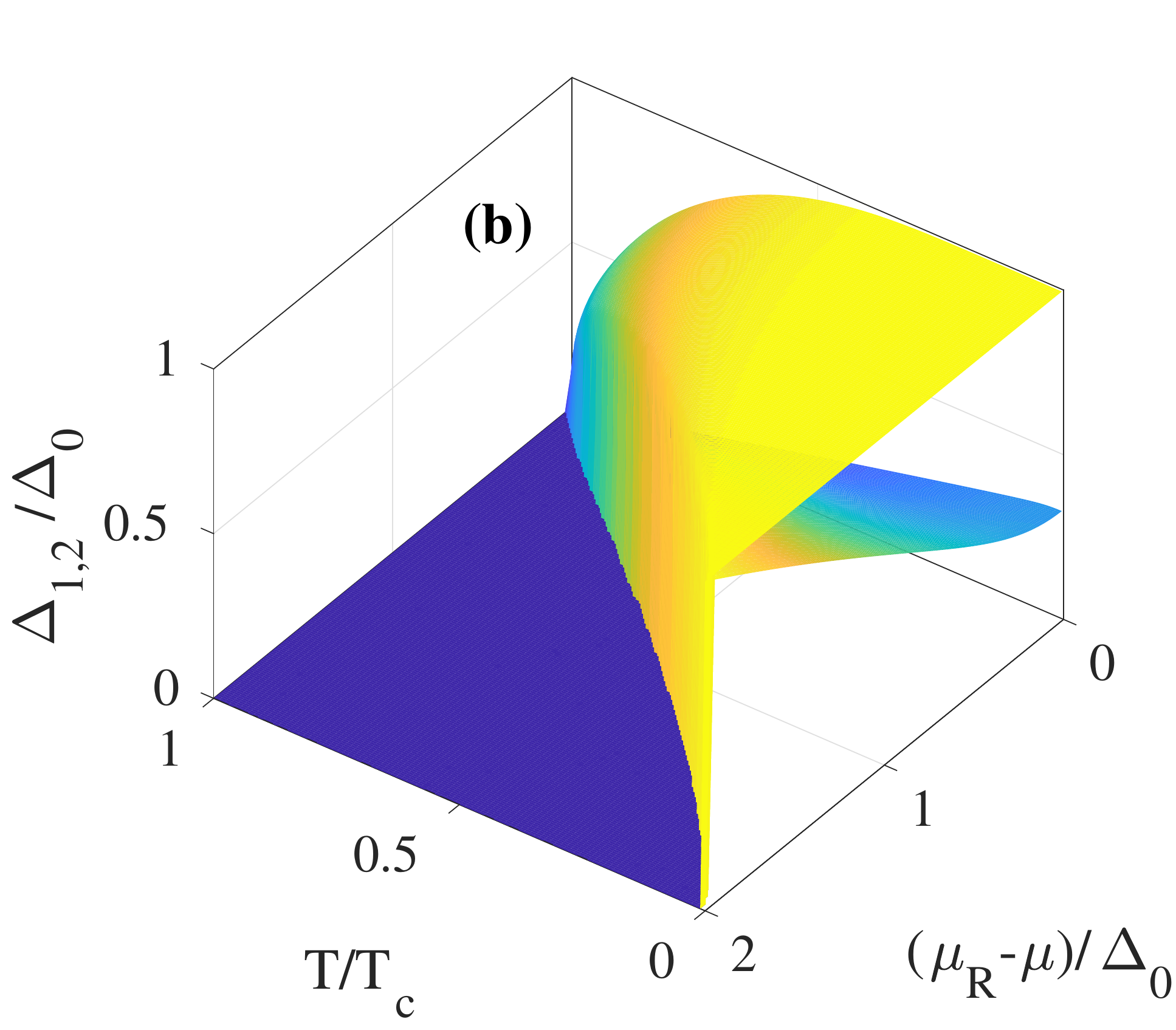}
  % Delta_1_2_vs_T_mu_a_v2.eps: 2320x2008 px, 300dpi, 19.64x17.00 cm, bb=0 0 557 482
  \caption{The two solutions for the energy gap. In (a) and (b) is presented the same figure, viewed from two different angles. In (b) we can see that the critical temperature is the same for both solutions.}
  \label{Delta_1_2_vs_T_mu_a_v2}
\end{figure}

As mentioned above, the Eqs.~(\ref{set_eqs2}) admit two sets of solutions $\Delta_{1,2}(\mu_R-\mu, T)$,  depicted in Fig.~\ref{Delta_1_2_vs_T_mu_a_v2}, corresponding to two functions $F_{1,2}(\mu_R-\mu, T)$.
In Figs.~\ref{Delta_1_2_vs_T_mu_a_v2}(a) and (b) are presented the same  functions, viewed from different perspectives.
The solutions with higher values, say $\Delta_1(\mu_R-\mu, T)$, were obtained also in Ref.~\cite{PhysicaA.464.74.2016.Anghel}, whereas the limits $\lim_{T = 0}\Delta_{1,2}(\mu_R-\mu, T)$ were found in~\cite{PhysicaA.531.121804.2019.Anghel}.
The solution $\Delta_{1}(\mu_R-\mu = 0, T)$ is the standard BCS solution (symmetric AB), with $\Delta_{1}(\mu_R-\mu = 0, T = 0) \equiv \Delta_0$, whereas $\Delta_2(\mu_R-\mu = 0, T = 0) \equiv \Delta_0/3$~\cite{PhysicaA.531.121804.2019.Anghel}.
In Fig.~\ref{Delta_1_2_vs_T_mu_a_v2}(b) we may observe that the energy gap for both solutions vanish at the phase transition temperature $T_{ph}(\mu_R - \mu)$,  where the standard BCS critical temperature is $T_c \equiv T_{ph}(\mu_R - \mu = 0)$.

\begin{figure}[t]
  \centering
  \includegraphics[width=6cm,bb=0 0 557 482,keepaspectratio=true]{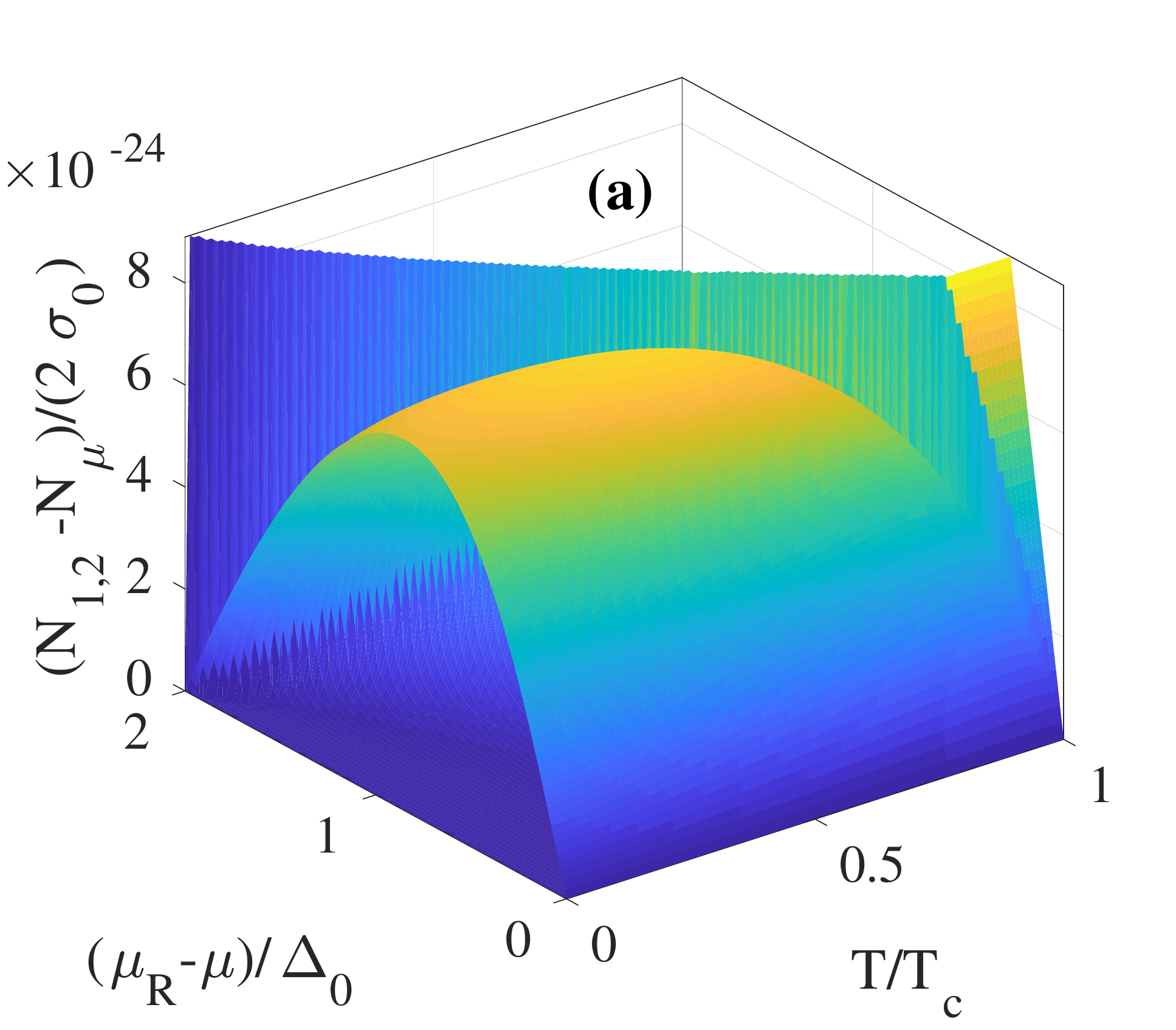}
  \includegraphics[width=6cm,bb=0 0 557 482,keepaspectratio=true]{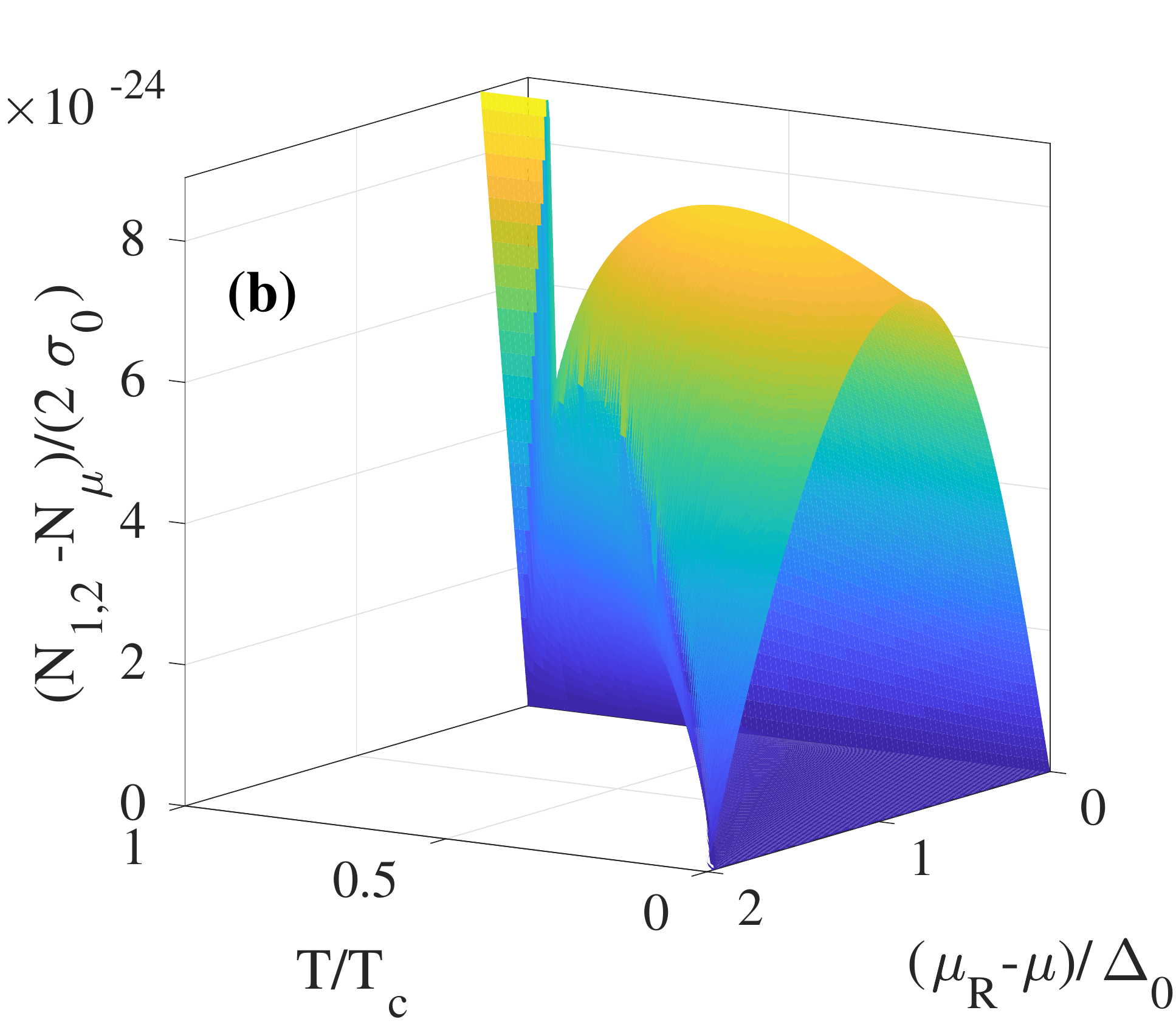}
  % Ntot_1_2_vs_T_mu_a_v2.eps: 2320x2008 px, 300dpi, 19.64x17.00 cm, bb=0 0 557 482
  \caption{The total number of particles in the system for the two solutions of Eqs.~\ref{set_eqs2}.
  The same picture is presented from two different angles, to emphasize both the variation of the  number of particles and the phase transition temperature.
%   Above the critical temperature, $N_{1,2}(\mu_R-\mu, T \ge T_c) - N_\mu = 2 \sigma_0 (\mu_R-\mu)$ and therefore has a jump for any $\mu_R-\mu > 0$, whcih can be seen in (a)
  At the phase transitrion $N_{1,2}(\mu_R-\mu, T \ge T_c) - N_\mu$ has a jump, which can be seen in (a).
  The jump was partly removed in (b) because it was masking some parts of the plot.
  }
  \label{Ntot_1_2_vs_T_mu_a_v2}
\end{figure}

In Fig.~\ref{Ntot_1_2_vs_T_mu_a_v2} we plot the total number of particles in the system (minus the number of single-particle states up to $\epsilon^{(0)} = \mu$) $N_{1,2}(\mu_R-\mu, T) - N_\mu$, corresponding to the two solutions of Eqs.~(\ref{set_eqs2}).
If the density of states is constant, below the phase transition temperature $T_{ph}(\mu_R-\mu)$ we have~\cite{PhysicaA.464.74.2016.Anghel}
\begin{subequations} \label{N_Nmu_int_sconst2}
\begin{eqnarray}
  &&
  N_{1,2}(\mu_R-\mu, T) - N_\mu = 2 \sigma_0 \int_{0}^{\hbar\omega_c} ( n_{\xi} - n_{-\xi}) \frac{\xi}{\epsilon} d\xi \nonumber \\
  &&
  = 2 \sigma_0 \int_{\Delta}^{\hbar\omega_c} ( n_{\sqrt{\epsilon^2 - \Delta^2}} - n_{-\sqrt{\epsilon^2 - \Delta^2}}) d\epsilon ,
  \label{N_Nmu_int_sconst2_e1}
\end{eqnarray}
whereas above $T_{ph}(\mu_R-\mu)$ we have
\begin{equation}
  N_{1,2}(\mu_R-\mu, T) - N_\mu = 2\sigma_0 (\mu_R - \mu) . \label{N_Nmu_int_sconst2_e2}
\end{equation}
\end{subequations}
From Eqs.~(\ref{N_Nmu_int_sconst2}) we observe that at phase transition $N_{1,2}(\mu_R-\mu, T_{ph}) - N_\mu$ has a jump, if $\mu_R \ne \mu$.
This can be observed in Fig.~\ref{Ntot_1_2_vs_T_mu_a_v2}(a) as a vertical surface behind the part of the plot which corresponds to $T \le T_{ph}$.

\section{Discussion} \label{sec_discussion}

Eventually the most interesting aspect that we can observe is that Eqs.~(\ref{set_eqs2}) have two solutions even in the limit $\mu_R \to \mu$ and these solutions correspond to well defined physical quantities, as we can see in Figs.~\ref{Delta_1_2_vs_T_mu_a_v2} and \ref{Ntot_1_2_vs_T_mu_a_v2}.
Apparently, this is in contradiction with the fact that, when $\mu_R - \mu \to 0$, the term $(\mu_R-\mu)(\xi - F)$ from the expression (\ref{pop_til_eps_sigma0}) of $n_{\xi}$ should vanish.
But this is not the case, because, as we shall see, $F_2$ (\ref{def_F_sigma0}) diverges in this limit.
If we introduce the notations
\begin{subequations} \label{defs_MM0}
\begin{equation}
  M_{1,2} (\mu_R - \mu, T) \equiv (\mu_R - \mu) F_{1,2}(\mu_R - \mu, T)
  \label{def_M}
\end{equation}
and
\begin{equation}
  M_{1,2}^{(0)}(T) \equiv \lim_{\mu_R \to \mu} M_{1,2} (\mu_R - \mu, T) ,
  \label{def_M0}
\end{equation}
\end{subequations}
we observe from Eq.~(\ref{pop_til_eps_sigma0}) that
\begin{equation}
  \lim_{\mu_R \to \mu} n_{\xi} = \frac{1}{e^{\beta(\epsilon_{\xi} + M_{1,2}^{(0)}/\epsilon_\xi)}+1}
  = \lim_{\mu_R \to \mu} n_{- \xi} . \label{pop_til_eps_sym}
\end{equation}
If we denote the numerator and  the denominator of $F$ (\ref{def_F_sigma0}) by
\begin{subequations} \label{defs_Fu_Fd}
\begin{eqnarray}
  F_{n\,1,2}(\mu_R-\mu, T) &\equiv& \int_{-\hbar\omega_c}^{\hbar\omega_c} \frac{( 1 - 2 n_{\xi} )\xi}{\epsilon^3} \, d\xi 
%   = 2 \left( \int\limits_\Delta^{\sqrt{(\hbar\omega_{c1})^2 + \Delta^2}} \frac{n_{-\xi(\epsilon)}}{\epsilon^2} d\epsilon
%   - \int\limits_\Delta^{\sqrt{(\hbar\omega_{c2})^2 + \Delta^2}} \frac{n_{\xi(\epsilon)}}{\epsilon^2} d\epsilon \right)
  \label{def_Fu}
  \qquad {\rm and} \\
  F_{d\, 1,2}(\mu_R-\mu, T) &\equiv& \int_{-\hbar\omega_c}^{\hbar\omega_c} \frac{(1 - 2 n_{\xi})}{\epsilon^3} d\xi
%   \approx \frac{2}{\Delta^2}
%   - 2 \left( \int\limits_\Delta^{\sqrt{(\hbar\omega_{c1})^2 + \Delta^2}} \frac{n_{-\xi(\epsilon)} d\epsilon}{\epsilon^2 \sqrt{\epsilon^2 - \Delta^2}}
%   + \int\limits_\Delta^{\sqrt{(\hbar\omega_{c2})^2 + \Delta^2}} \frac{n_{\xi(\epsilon)} d\epsilon}{\epsilon^2 \sqrt{\epsilon^2 - \Delta^2}}\right)
  \label{def_Fd}
\end{eqnarray}
\end{subequations}
we further observe that, if $M_i^{(0)}(T)$ is different from zero (where $i=1$ or $2$), then the condition $\lim_{\mu_R\to\mu} |F_i(\mu_R-\mu, T)| = \infty$ must be satisfied, which implies $\lim_{\mu_R\to\mu} F_{d \, i}(\mu_R-\mu, T) = 0$.
But we also notice that $\lim_{\mu_R\to\mu} F_{n\, i}(\mu_R-\mu, T) = 0$ as well, due to the symmetry of $n_\xi$ in $\xi$, in the limit $\mu_R \to \mu$ (Eq.~\ref{pop_til_eps_sym}).
Nevertheless, the limit $M^{(0)}_i(T)$ must remain different from zero for any $T < T_c$.

\begin{figure}[t]
  \centering
  \includegraphics[width=6cm]{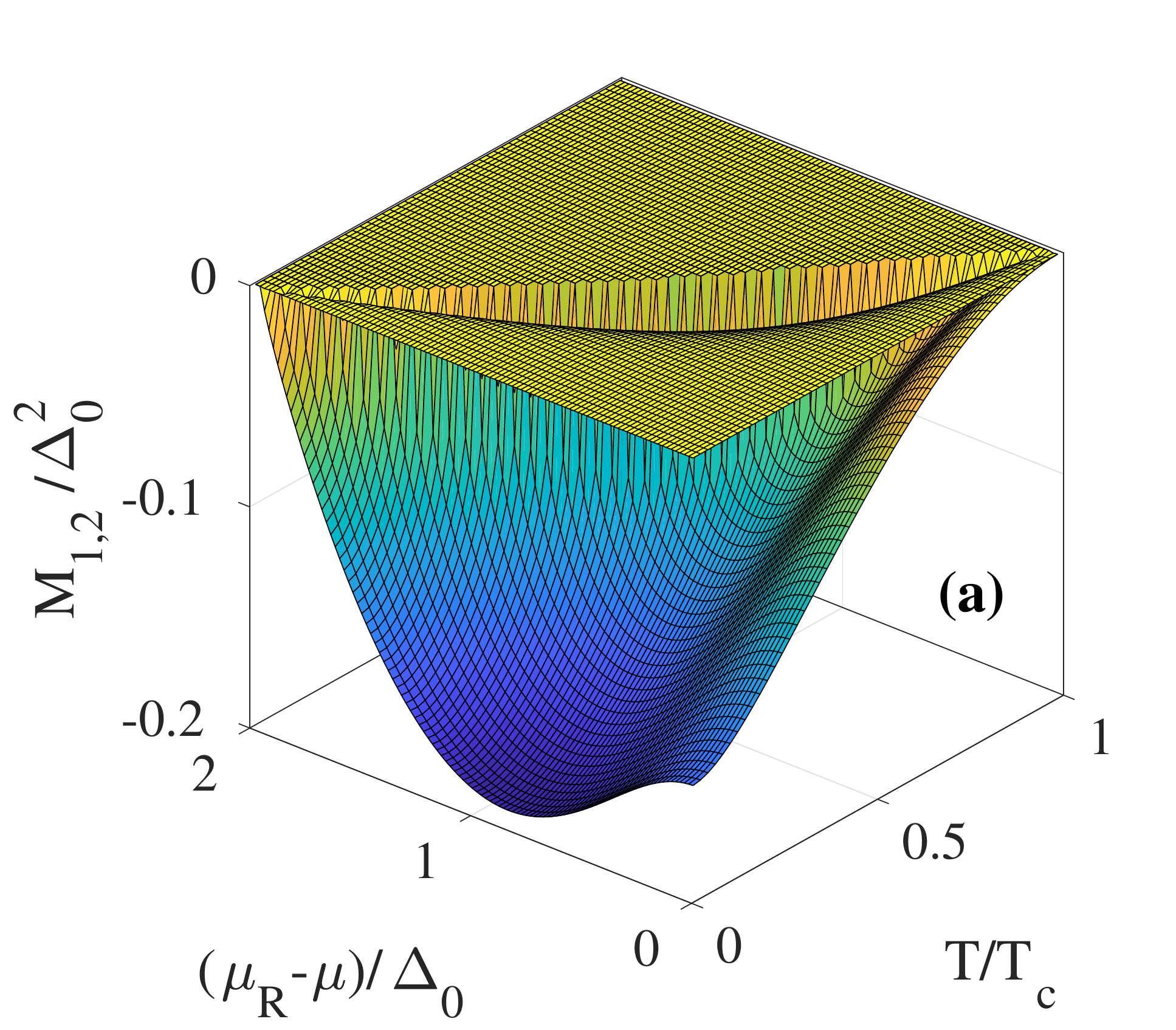}
  \includegraphics[width=6cm]{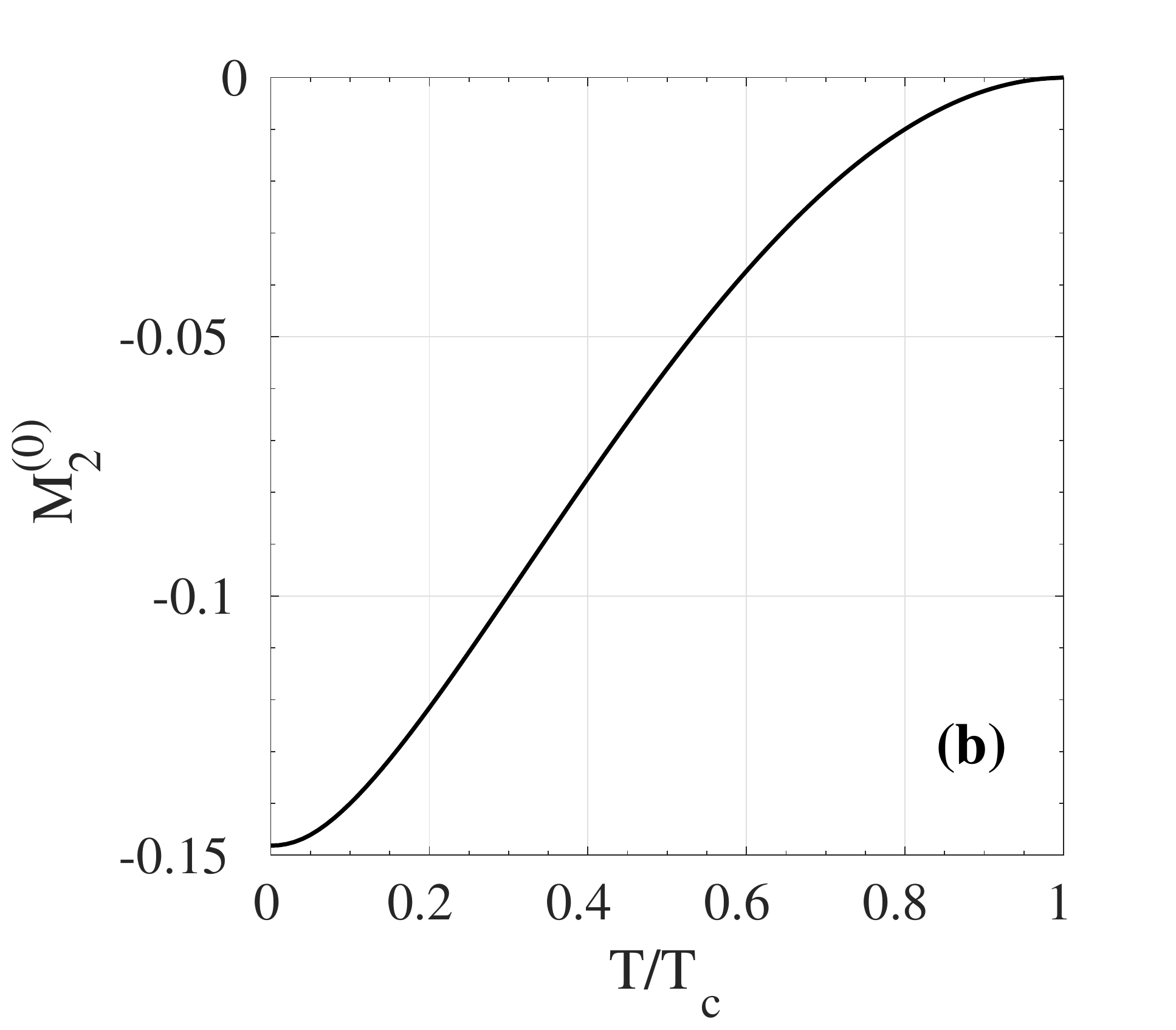}
  % M0_vs_T.eps: 2320x2033 px, 300dpi, 19.64x17.21 cm, bb=0 0 557 488
  \caption{(a) $M_{1,2} \equiv (\mu_R-\mu) F_{1,2}$ and (b) $M^{(0)}_2 \equiv \lim_{\mu_R\to\mu} M_2$ ($\lim_{\mu_R\to\mu} M_1 = 0$).}
  \label{M0_vs_T}
\end{figure}

In  Fig.~\ref{M0_vs_T}(a) we plot $M_{1,2}(\mu_R-\mu, T)$.
We observe that $\lim_{\mu_R\to\mu} M_1(\mu_R-\mu, T) = 0$ and this case corresponds to the standard BCS theory.
In this situation, Eqs,~(\ref{set_eqs2}) become the standard BCS equations for the energy gap and populations.
In Fig.~\ref{Delta_1_2_vs_T_mu_a_v2}(a) we can see the  solution for the energy gap, $\Delta_1$, for $\mu_R-\mu = 0$.

From the second function plotted in Fig.~\ref{M0_vs_T}(a), namely $M_2(\mu_R-\mu, T)$, we see that $M_2^{(0)}(T)$ is different from zero for any $T < T_c$ (Fig.~\ref{M0_vs_T}b).
Using the results of Ref.~\cite{PhysicaA.531.121804.2019.Anghel} we can readily calculate $M_2^{(0)}(0) = -(7/27) \Delta_0^2$.
The curve $M_2^{(0)}(T)$ may be either calculated as the limit of $M_2(\mu_R-\mu, T)$, as $\mu_R$ converges to $\mu$ (as it was done in Fig.~\ref{M0_vs_T}b) or by solving the self-consistent set formed by the Eqs~(\ref{Eq_int_Delta1}), (\ref{pop_til_eps_sym}),  and $F_{d\, 2}(0, T) = 0$ (from Eq.~\ref{def_Fd}).
% We observe that in this case, the quasiparticle populations are not zero and they obey the ``balance'' equation $F_{d\, 2}(0, T) = 0$.
At zero temperature, the last equation leads to the total occupation of the quasiparticle states in the interval $\xi \in [- \Delta_0/(3\sqrt{3}), \Delta_0/(3\sqrt{3})]$ (see Ref.~\cite{PhysicaA.531.121804.2019.Anghel}).
Furthermore, the situation is symmetric with respect to $\mu$, since for $\mu_R-\mu \le 0$, $F \ge 0$, %the energy gap is the same, whereas
and the quasiparticle levels populations are mirrored with respect to $\mu$ (that is $n_\xi$ becomes $n_{-\xi}$)~\cite{PhysicaA.464.74.2016.Anghel}.

From the point of view of the partition function, the existence of the second solution is quite easy to understand as well.
The partition function and the condition of equilibrium are~\cite{PhysicaA.464.74.2016.Anghel}
\begin{eqnarray}
  \ln(\cZ)_{\beta\mu} &=& - \sum_{\xi, i} [(1 - n_{\xi i}) \ln(1 - n_{\xi i}) + n_{\xi i} \ln n_{\xi i} ] \nonumber \\
  && - \beta (E-\mu_R N) \label{cZ_beta_mu}
\end{eqnarray}
and
\begin{equation}
  \frac{\partial\ln(\cZ)_{\beta\mu}}{\partial n_{\xi i}} = \ln\frac{1-n_{\xi i}}{n_{\xi i}} - \beta \left[ \epsilon_\xi - \frac{\mu_R - \mu}{\epsilon_\xi} (\xi - F)\right] = 0, \label{dlnZ_dnki}
\end{equation}
respectively (again, $i = 0,1$) -- without knowing of the second solution, presented here and in Ref.~\cite{PhysicaA.531.121804.2019.Anghel}, in Ref.~\cite{PhysicaA.464.74.2016.Anghel} it was stated that when $\mu_R = \mu$, the standard BCS theory is restored, which is only partially true, as we have seen above.
If $F$ diverges when $\mu_R \to \mu$, in such a way that $M$ (given by Eqs.~\ref{defs_MM0}) converges to a non-zero value, then indeed, the quasiparticle population used above (Eqs.~\ref{pop_til_eps_sigma0} and \ref{pop_til_eps_sym}) maximize the partition function,  according to Eq.~(\ref{dlnZ_dnki}), and therefore represent equilibrium distributions.
We may also notice that the divergence of $F$ is due to the divergence of the derivative of the energy gap with respect to the population of any quasiparticle level~\cite{PhysicaA.464.74.2016.Anghel}
\begin{equation}
  \frac{\partial\Delta}{\partial n_{\xi i}} = - \left\{\sigma_0 \Delta \epsilon \left[ \int_{-\hbar\omega_c}^{\hbar\omega_c} \frac{(1 - n_{\xi 0} - n_{\xi 1}) d\xi}{\epsilon^3} \right] \right\}^{-1} \label{dDelta_dNxi}
\end{equation}
(in Eq.~\ref{dDelta_dNxi} I intentionally left both, $n_{\xi 0}$ and $n_{\xi 1}$, since this equation does not refer only to the equilibrium distribution, where $_{\xi 0} = n_{\xi 1}$).
In other words, for the second (the non-standard) solution, when $\mu_R-\mu \to 0$, the sensitivity of the energy gap $\Delta$ with respect to the population variation (that is, $\partial\Delta/\partial n_{\xi i}$) diverges.
This divergence is manifested in both, the variation of the total particle number and of the total energy of the system and leads to a finite contribution to the equilibrium quasiparticle populations.

\section{Conclusions} \label{sec_conclusions}

In conclusion, in this paper I analysed the solutions of the energy gap equation in the BCS formalism by diagonalizing the model Hamiltonian $\hat \cH - \mu \hat N$, where $\mu$ is the center of the attraction band (AB) and may be different from the chemical potential $\mu_R$.
The solutions are symmetric with respect to the change of sign $\mu_R - \mu \to - (\mu_R - \mu)$~\cite{PhysicaA.464.74.2016.Anghel}, so I made the calculations only the case $\mu_R - \mu \ge 0$.
Therefore, if $|\mu_R - \mu| < 2 \Delta_0$ (where $\Delta_0$ is the energy gap in the standard BCS theory, at zero temperature), the superconducting phase is formed below a phase transition temperature $T_{ph}(\mu_R-\mu)$ which decreases with $|\mu_R - \mu|$, reaching $T_{ph}(\mu_R - \mu = \pm 2\Delta_0) = 0$.
For each value of the difference $|\mu_R - \mu| < 2\Delta_0$ and for temperatures below $T_{ph}(\mu_R-\mu)$ there are two solutions for the energy gap, $\Delta_1(\mu_R-\mu, T)$ and $\Delta_2(\mu_R-\mu, T)$, corresponding to two distinct quasiparticle distributions.
Both solutions have the same phase transition temperature (as it is implied by the notation $T_{ph}$ above).
Eventually the most interesting result is that event at $\mu_R = \mu$ -- which is the standard and the most simple BCS case -- there are still two solutions for the energy gap and two particle distributions that satisfy the equilibrium condition -- which is the maximization of the partition function.
One of these solutions is the standard one, with the energy gap, $\Delta_1(\mu_R - \mu = 0, T)$, where $\Delta_1(\mu_R - \mu = 0, T=0) = \Delta_0$.
The second solution has a smaller energy gap, $\Delta_2(\mu_R - \mu = 0, T) < \Delta_1(\mu_R - \mu = 0, T)$ (for any $T < T_{ph}(\mu_R-\mu = 0)$) and the phase  transition temperature is the critical temperature of the BCS theory, i.e. $T_{ph}(\mu_R-\mu = 0) = T_c$.

This work has been financially supported by UEFISCDI project PN-19060101/2019. Travel support from Romania-JINR collaboration projects positions 23, 26, Order 397/27.05.2019 is gratefully acknowledged.

%\section*{References}

\end{document}